\documentclass[runningheads]{llncs}
\usepackage{lineno}
\modulolinenumbers[5]
\usepackage[utf8]{inputenc}
\usepackage[T1]{fontenc}
\usepackage{cite}
\usepackage{amsmath,amssymb,amsfonts,mathrsfs}
\usepackage{algorithmic}
\usepackage{graphicx}
\usepackage{array}
\usepackage[caption=false,font=footnotesize,labelfont=sf,textfont=sf]{subfig}
\usepackage{textcomp}
\usepackage[table,dvipsnames]{xcolor}
\usepackage{booktabs} 
\usepackage{multirow}
\usepackage{listings}
\def\BibTeX{{\rm B\kern-.05em{\sc i\kern-.025em b}\kern-.08em
    T\kern-.1667em\lower.7ex\hbox{E}\kern-.125emX}}

\usepackage{xspace}
\newcommand{\ie}{i.e.,\xspace}
\newcommand{\eg}{e.g.,\xspace}

\xspaceaddexceptions{=}

\usepackage[acronym,toc,nonumberlist]{glossaries}
\makenoidxglossaries
\newacronym{fft}{FFT}{Fast Fourier Transform}

\usepackage{listings} 
\definecolor{col1}{HTML}{1E88E5}
\definecolor{col2}{HTML}{D81B60}
\definecolor{col3}{HTML}{43A047}
\definecolor{col4}{HTML}{F4511E}
\definecolor{col5}{HTML}{205B74}

\usepackage{pifont}

\usepackage[colorlinks=true, allcolors=blue, bookmarks=false]{hyperref}

\RequirePackage[normalem]{ulem}
\RequirePackage{color}\definecolor{RED}{rgb}{1,0,0}\definecolor{BLUE}{rgb}{0,0,1}

\providecommand{\DIFdel}[1]{{\protect\color{RED}\sout{#1}}}
\providecommand{\DIFdel}[1]{}

\begin{document}
%
\title{Software Development Vehicles to enable extended and early co-design: \\ a RISC-V and HPC case of study }
%
\titlerunning{RISC-V and HPC: enabling co-design with SDV}
%
\author{Filippo Mantovani\inst{1}\and 
Pablo Vizcaino\inst{1}\and 
Fabio Banchelli\inst{1}\and 
Marta Garcia-Gasulla\inst{1}\and 
Roger Ferrer\inst{1}\and 
Giorgos Ieronymakis\inst{2}\and 
Nikos Dimou\inst{2}\and 
Vassilis Papaefstathiou\inst{2}\and 
Jesus Labarta\inst{1} 
}
\authorrunning{F. Mantovani et al.}
%
\institute{Barcelona Supercomputing Center
Plaça Eusebi Güell, 1-3
08034 Barcelona (Spain)
\email{name.surname@bsc.es}\and
FORTH-ICS
N. Plastira 100, Vassilika Vouton,
GR-70013, Heraklion, Crete, Greece
\email{\{ieronym,ndimou,papaef\}@ics.forth.gr}
}
%
\maketitle              
\begin{abstract}


Prototyping HPC systems with low-to-mid technology readiness level (TRL) systems is critical for providing feedback to hardware designers, the system software team (\eg compiler developers), and early adopters from the scientific community. 
The typical approach to hardware design and HPC system prototyping often limits feedback or only allows it at a late stage.
In this paper, we present a set of tools for co-designing HPC systems, called software development vehicles (SDV).
We use an innovative RISC-V design as a demonstrator, which includes a scalar CPU and a vector processing unit capable of operating large vectors up to 16~kbits.
We provide an incremental methodology and early tangible evidence of the co-design process that provide feedback to improve both architecture and system software at a very early stage of system development.

\end{abstract}
\section{Introduction and related work}\label{secIntro}

%
%
%
%

\begin{figure}[!htbp]
\centering
\includegraphics[width=\columnwidth]{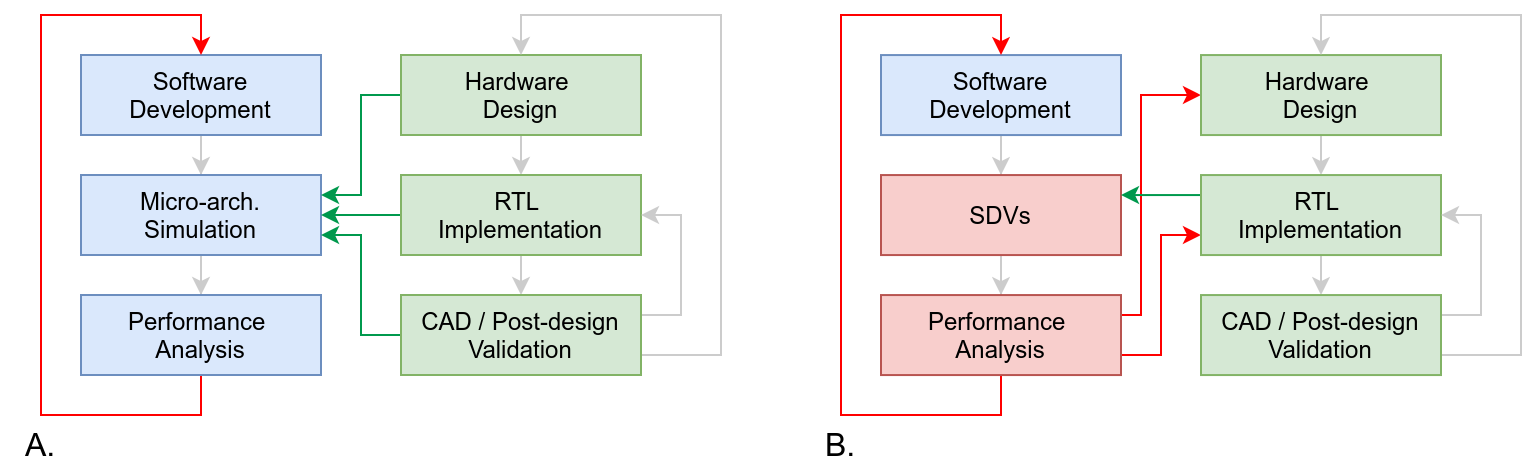}
\caption{Co-design flow for hardware (green) and software (blue)}
\label{figHsSwCodesign}
\end{figure}

The typical high-level approach to hardware design foresees to develop a new design, implementing it at Register Transfer Level (RTL) using Hardware Description Language (HDL), and finally maps the implementation on a given technology using Computer-Aided Design (CAD) tools for post-design validation. 
This flow is represented on the right part of Figure~\ref{figHsSwCodesign}A.

The software development for new hardware designs relies on a micro-architec\-tural simulator which collects the inputs from the hardware design and its implementation to mimic the behaviour of the proposed new hardware.
Software developers can therefore test their codes and analyze their performance thanks to the simulator.
This flow is represented on the left part of Figure~\ref{figHsSwCodesign}A.

Besides the fact that booting an operating system and running complex codes through a simulator can be extremely time consuming (or even impossible), from Figure~\ref{figHsSwCodesign}A it is clear that the software development (blue) can not influence much the hardware flow (green).

In this paper, we propose a methodology that allows software developers to provide feedback both to the architects designing the architecture and to the engineers implementing the RTL.
Moreover, our infrastructure guarantees the possibility of porting, testing, benchmarking, and optimizing software on the new proposed hardware as early as possible.
The proposed method is depicted in Figure~\ref{figHsSwCodesign}B. Instead of a software simulator, it leverages a collection of hardware platforms (mostly FPGAs) and software tools called Software Development Vehicles (SDVs).
The SDV allows software developers to test and analyze their codes on an environment using software tools for collecting insights from the executions while running on the latest RTL implementation of the proposed architecture.
Thanks to this infrastructure, software developers can therefore provide valuable feedback to the hardware team as represented by the red arrows of Figure~\ref{figHsSwCodesign}B.
Also, the SDV reduces the dependencies from the hardware design flow (green arrows), since it only depends on the RTL implementation of the new proposed architecture.

The method is conceptually similar to the one enabled with tools such as \eg Siemens Veloce\footnote{\url{https://eda.sw.siemens.com/en-US/ic/veloce/}}.
However, it is more lean, since it is working on a single FPGA (at the additional price of not having visibility on all signals of the system).
The method does not require software simulators, such as riscvOVPsim\footnote{\url{https://github.com/riscv-ovpsim/imperas-riscv-tests}} and FireSim~\cite{farshchi2019integrating} nor meta-hardware description languages, such as Chisel or SystemC~\cite{kim2019fpga}: on the positive side the whole SDV infrastructure is faster and allows to run OS, libraries and applications as in high TRL system. On the negative side, some modifications can require more time to be implemented, since they can involve RTL development.

For our evaluation, we have chosen to focus on the hardware development of a RISC-V-based design that targets the HPC domain and is developed within the European Processor Initiative project%
\footnote{https://www.european-processor-initiative.eu/}.
This design includes a RISC-V micro-tile, which is composed of an Avispado scalar core developed by Semidynamics%
\footnote{Semidynamics. \url{https://semidynamics.com/}},
connected to a Vitruvius vector processing unit (VPU)~\cite{minervini2023vitruvius}
with eight lanes.
Each lane has a Floating Point Unit (FPU) developed by the University of Zagreb~\cite{kovavc2023faust}.
The micro-tile also has a Home Node and an L2 cache, which were respectively designed by 
Chalmers%
\footnote{Chalmers University of Technology. \url{https://www.chalmers.se}} 
and FORTH%
\footnote{FORTH Institute of Computer Sciences. \url{https://www.ics.forth.gr/carv}}.
The micro-tile is connected to a few ancillary FPGA blocks in order to interface with on-board main memory (DDR4), the PCIe bus, and the Ethernet PHY. The most disruptive feature of this design is the presence of a vector processing unit that is capable of operating on vectors of 256 double-precision elements (\ie 16 kbits-wide vector registers).


The design point of such a RISC-V system is rather extreme, as it can operate on vectors that are up to 32 times larger than current SIMD architectures used in HPC\footnote{Intel x86 AVX512 operates SIMD vectors of 8 double-precision elements}. Therefore, it is crucial for the software community to have an environment where to test the behavior of current scientific applications and the readiness of system software to exploit such a design. At the same time, it is important for RTL designers and system software developers to receive feedback from application developers so that they can improve the efficiency of their implementation while still keeping it as general-purpose as possible. We consider this paper a contribution to the HPC community and an example of co-design since the Software Development Vehicles tools introduced improve the design cycle of HPC hardware and software.


The main contributions of this paper are:
{\em i)} the methodology for building a stronger connection between software and hardware during its design phase;
{\em ii)} the tools and the infrastructure needed to run a complex software stack on top of implementation of early hardware designs;
{\em iii)} the evaluation of the proposed methodology with a library used for the computation of the Fast Fourier Transform
running on an emerging design that includes a RISC-V core and a vector processing unit (VPU).

The rest of this paper is structured as follows:
Section~\ref{secSdvComponents} lists the hardware and components of the SDV environment.
Section~\ref{secMethodology} explains the steps that compose the proposed methodology.
In Section~\ref{secEvaluation} we provide evidences of the methodology, applying it to a code for computing Fast Fourier Transform on a RISC-V vector architecture.
We conclude the paper with Section~\ref{secConclusions} where we summarize some conclusions remarks.

\section{Components of the Software Development Vehicles (SDVs)}\label{secSdvComponents}

\subsection{RISC-V commercial platforms}\label{secCommercialPlatforms}


There are several RISC-V based commercial platforms on the market.
The compute node of our cluster is powered by a RISC-V-based SoC by SiFive (Freedom U740 SoC).
Each SoC houses four 64-bit RISC-V cores running at up to 1.2~GHz.
Each core supports the I, M, A, F, D, and C extensions of the RISC-V Instruction Set Architecture (ISA).
The SiFive SoC is mounted on a miniITX PCB with 4~GB of RAM, a slot for a micro-SD for the boot of the OS and an m.2 connector for SSD where we store the filesystem.
We assembled two motherboards in a 1U chassis, so to have a higher density, similar to the system described in~\cite{montecimone}.
Nodes of the cluster are connected using the on-board 1~GbE Ethernet link.
The board is known on the market as HiFive Unmatched and more details can be found on the provider web site: 
\url{https://www.sifive.com/boards/hifive-unmatched}.
The nodes of the cluster are operated with a standard Linux distribution (Ubuntu 20.04 at the moment of the writing of this document).
A standard GNU Compiler Suite is available on those platforms. 
However, BSC developed an LLVM-based compiler toolchain which supports vector specifications v0.7.1 and v1.0.
Vectorization can be achieved
{\em i)} enabling compiler auto-vectorization capabilities,
{\em ii)} adding vector pragmas, or
{\em ii)} manually using intrinsics.

\subsection{Software emulation: Vehave}


Vehave is a user-space emulator for the vector extension of the RISC-V ISA (RVV) that runs on RISC-V Linux. 
It allows a functional verification of a program that uses RVV instructions or a code generator, such as a compiler, that emits RVV instructions. 
It runs on top of the RISC-V commercial platforms described in Section~\ref{secCommercialPlatforms} (configured by the user running the {\tt module} environment) and it is distributed as binary here 
\url{https://ssh.hca.bsc.es/epi/ftp/vehave-EPI-development-latest.tar.bz2}.

Vehave emulates instructions by intercepting the illegal instruction exception that a CPU emits when it encounters an unknown/invalid instruction. 
Once an illegal instruction is found, Vehave decodes it and if it is a valid RVV instruction it emulates it, else an error is propagated back.
The program resumes once the emulation of the vector instruction is complete.
Vehave relies on the LLVM libraries of the compiler which already supports the RVV for the process of decoding the instructions.
The output of Vehave is collected in a {\tt .trace} file which stores in plain text extensive details about each vector instruction emulated and some quantitative figure of the scalar code executed before each vector instruction.
The {\tt .trace} file generated by Vehave is parsed and converted to {\tt .prv} format which can be visualized with Paraver~\cite{pillet1995paraver}.
These traces may not include cycle accurate data, 
but include detailed data on the vector instructions executed that are valuable to study regions of code with potential for vectorization.
%
%

\subsection{FPGA-based emulation}

%
The FPGA-based emulation platform comprises of an FPGA evaluation board and a host~x86 server.
%
%
The server is a commodity server housing an AMD Ryzen 5 5600 CPU with 32GB of DDR4-3200 memory, both mounted on a Mini-ITX motherboard.
It runs a regular Ubuntu Server 20.04 with local storage and a mounted Network Filesystem.
%

%
The FPGA board is the Virtex UltraScale+ HBM VCU128 FPGA Evaluation Kit\footnote{\url{https://www.xilinx.com/products/boards-and-kits/vcu128.html}}, commonly referred to in the rest of the document as {\em VCU128}.
The board includes a VU37P FPGA\footnote{Complete device name: XCVU37P-2FSVH2892E}, with 8~GB of integrated HBM memory.
There are also five on-board DDR4 memory modules, adding up to a total of 4.5~GB of memory.

\begin{figure}[htbp]
  \centering
  \includegraphics[width=.75\linewidth]{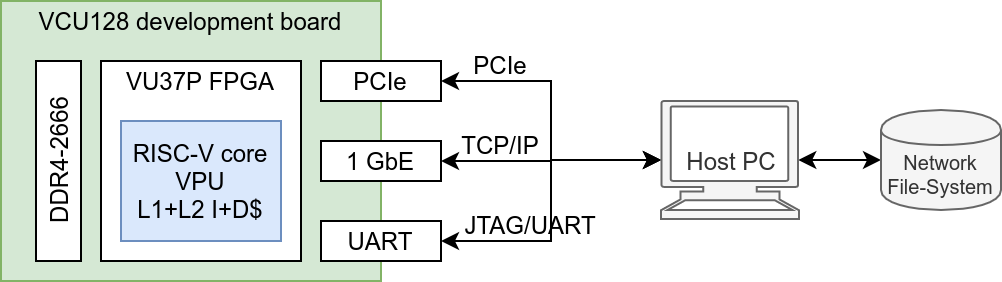}
  \caption{Connection between host server and VCU128 board: schematic view}
  \label{figSdvComponents}
\end{figure}
%
Figure~\ref{figSdvComponents} shows the connection scheme between the host server and the VCU128 device.
The host and the VCU128 are connected through three different interfaces.
The first one being the UART/JTAG interface, used to program the FPGA and also to access a UART terminal once a Linux image is running on the VCU128.
The Ethernet PHY connector establishes a point-to-point TCP/IP network between host and device that is used to access the RISC-V SoC running on the FPGA via SSH and give access to a network filesystem (NFS) to the RISC-V SoC once booted Linux.
Lastly, the VCU128 is configured to operate its PCIe interface using 16~lanes of Gen~3 PCIe. The PCIe link is used to pre-load the on-board DDR4 memory with the Linux image and configure the VCU128 board to autonomously boot Linux.
The RISC-V SoC runs the same Linux distribution that is deployed on the RISC-V commercial platforms described in Section~\ref{secCommercialPlatforms}, so that binaries (both statically or dynamically generated) are compatible on the two platforms.
A compute node is composed of the host server and the VCU128 board.
There are several compute nodes available via a job scheduler (SLURM) for users of the SDV platform.

\paragraph*{Hardware counters}
When running on native SoC, even on FPGA, hardware counters provide information about the micro-architecture.
These counters are available through CSR read instructions.
Since counters are protected and can only be read in machine mode, the SBI interface bridges between machine mode and system calls.
We implement a custom system call that bridges between user code and kernel space.

\paragraph*{Tracing executions}
The PAPI library~\cite{mucci1999papi} offers a standard interface for users to read hardware counters.
Our current implementation allows reading \texttt{hpmcounter{0-4}} and other contextual information (\eg current vector-length).
Users can manually instrument their code with calls to PAPI or use Extrae~\cite{extrae}.
Extrae is a tracing tool that polls hardware counters via manual instrumentation or with automatic hooks of supported libraries such as MPI and OpenMP.
%
%
%
We leverage the Integrated Logic Analyzer (ILA)\footnote{\url{https://www.xilinx.com/products/intellectual-property/ila.html}}, a signal-level monitor of Xilinx-based FPGAs,
to record the values of selected signals during each clock from a given time window.
The start time of the monitoring is triggered upon a user-defined condition (\eg a signal value must be zero).
%

\paragraph*{Visualization}
Paraver~\cite{pillet1995paraver} is a trace visualization tool developed at BSC.
In-house tools convert Vehave, Extrae, and ILA traces to a format that can be understood by Paraver.
Using the same visualization tool to analyze experiments in different platforms of the SDV ecosystem and with different levels of detail is key to {\em i)} achieve wide adoption across users and {\em ii)} accelerating the feedback loop between software and hardware teams.

\section{Co-design methodology}\label{secMethodology}

\begin{figure}[htbp]
  \centering
  \includegraphics[width=.8\linewidth]{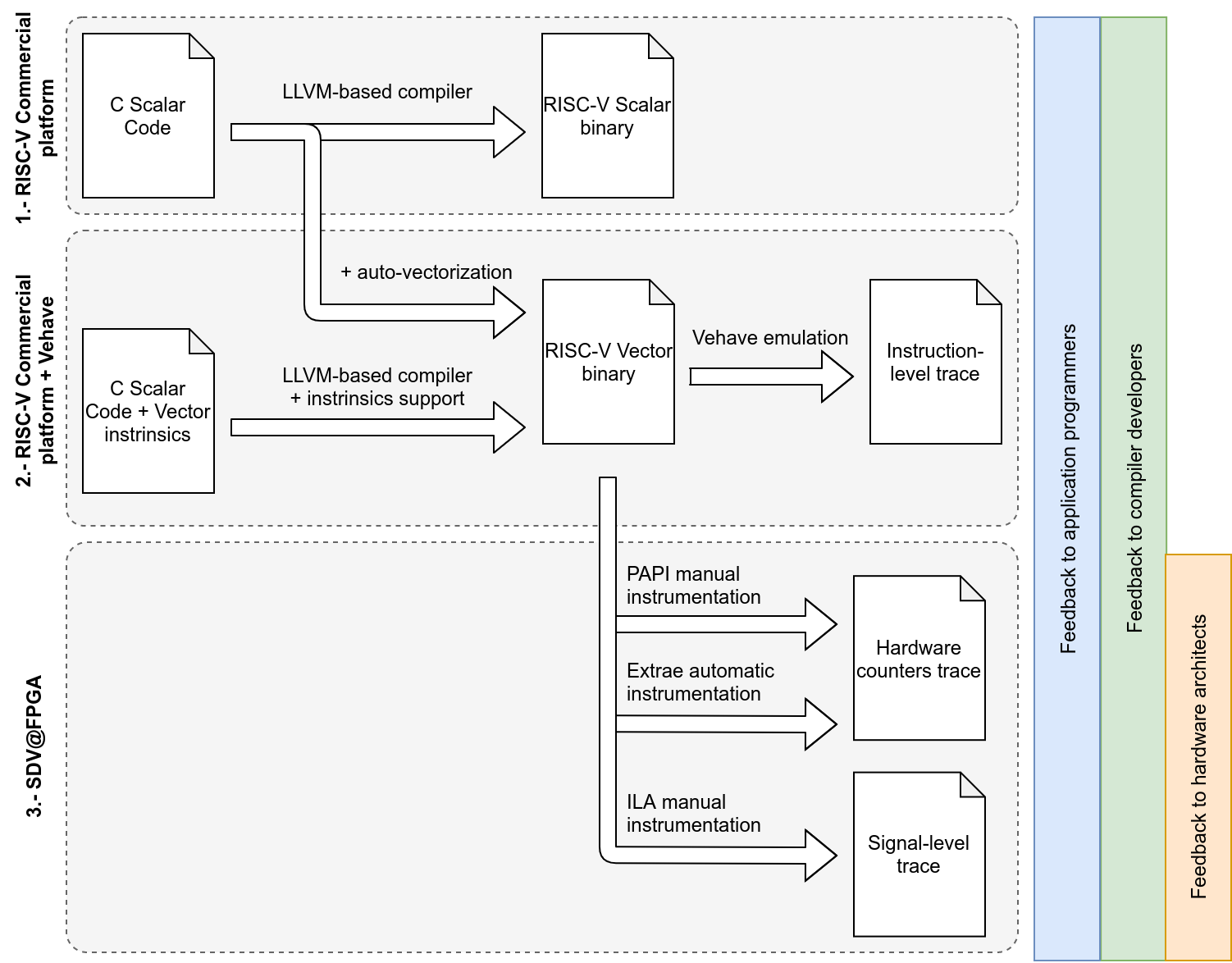}
  \caption{Performance analysis workflow in SDV} 
  \label{figSdvWorkflow}
\end{figure}

In this section we present the methodology that creates a feedback loop between hardware and software teams thanks to the SDV approach.
We propose an evaluation workflow with three steps as depicted in Figure~\ref{figSdvWorkflow}.

\subsection{Porting to scalar RISC-V commercial platforms}
First, users port their application to the RISC-V architecture.
The LLVM-based compiler generates binaries compatible with the \texttt{rv64gc} architecture.
At this stage, the generated binary is RISC-V compatible but it contains only scalar instructions.
The application can run natively, using the commercial RISC-V platforms.
This step is useful 
{\em i)} to verify the compatibility of the code under study with the RISC-V architecture (\eg no intrinsics or inline assembly of other architectures);
{\em ii)} to verify that the compiler supports all data structures and code features required by the code under study;
{\em iii)} to benchmark the code on a commercial scalar RISC-V.

\subsection{Vectorization and software emulation}
The next step is to vectorize the code, leveraging compiler auto-vectorization, using pragmas or adding vector intrinsics.
The resulting binary includes vector instructions. When this binary runs on the RISC-V commercial platforms together with Vehave, the vector instructions will be emulated and details about their execution stored in a trace file that can be analyzed post mortem.
In this step we can gather information about
{\em i)} potential of the code to be vectorized or patterns that can prevent its vectorization,
{\em ii)} ability and limitations of the compiler to auto-vectorize code,
{\em iii)} efficiency of the vectorization, \eg checking if the code exploits the optimal vector length.

The resulting binary includes vector instructions, which can be run through Vehave.
This step allows the user to validate the correctness of their code and discover the potential for vectorization by analyzing the instruction-level traces.
\subsection{Performance analysis on the FPGA prototype}
Lastly, users can now use the FPGA development platform, where the same vectorized binary runs on a RISC-V core with support for the RVV (vector) ISA extension.
The whole binary is run natively, with no software emulation.
During this phase, developers have access to hardware counters, which enables performance analysis at the micro-architecture level.
Information such as the number of cycles when the vector unit was active, or cycles lost due to pipeline stalls is available through a standard PAPI interface.
If the evaluation requires a finer grained analysis, a selected number of RTL signals can also be monitored during the execution of the application.
Developers can manually instrument their code to trigger the ILA at a certain execution point and conduct a signal-level analysis.

Section~\ref{secEvaluation} presents a case study of the evaluation methodology presented in this section.
Throughout the steps depicted in Figure~\ref{figSdvWorkflow}, users of SDV can provide feedback to other teams:
{\em i)} Instruction-level traces to study the algorithm implementation and give feedback to both the application programmers and the compiler developers.
{\em ii)} Hardware counters and signal-level traces to study the effects of micro-architectural features to give feedback to hardware architects.

\section{Evaluation}\label{secEvaluation}

To showcase the potential and benefits of the SDV, we use our proposed methodology to evaluate a vectorized \Gls{fft} implementation for RISC-V~\cite{fftp}.

\subsection{Step 1: porting to RISC-V scalar commercial platforms}
As presented in Section~\ref{secMethodology}, the first step proposed in the SDV methodology is running the application on a scalar RISC-V platform. 
Firstly we compiled the FFTW~\footnote{\url{https://www.fftw.org}} using the LLVM-based compiler, confirming that although the library could be run on RISC-V, it did not include vector instructions. 
Our next step was to code an FFT algorithm with the potential for long-vector vectorization, and again try it on the scalar platforms for verification purposes.
After this was done, we started vectorizing and analyzing the implementation.

\subsection{Step 2: vectorization and software emulation}

\begin{figure}[htbp]
  \centering
  \includegraphics[width=.8\linewidth]{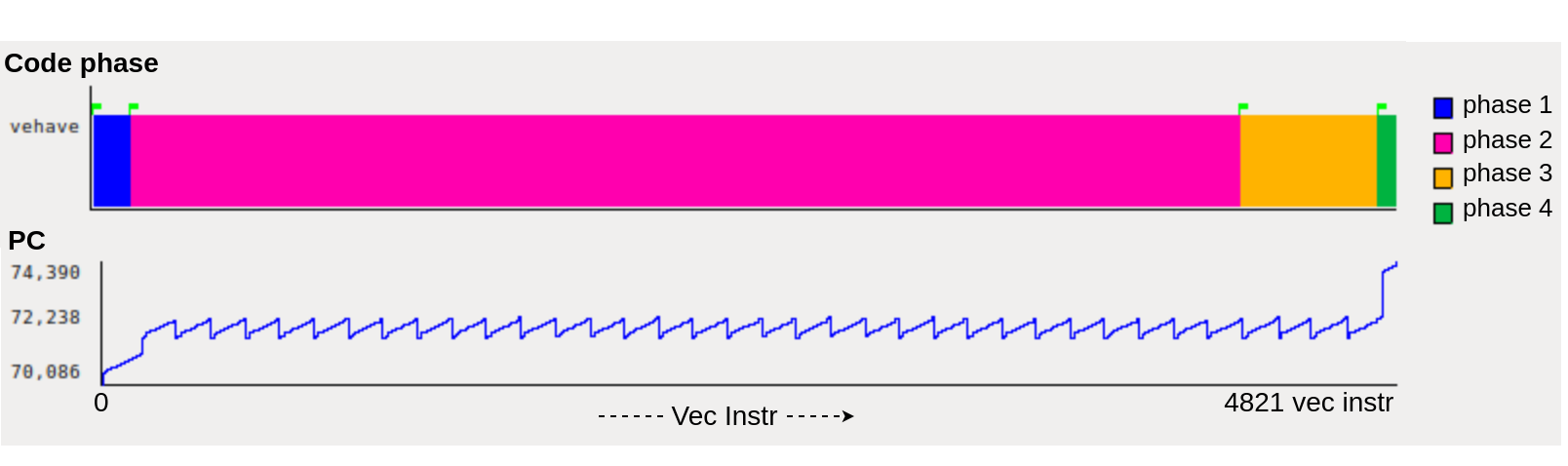}
  \caption{Vehave trace with Code phase (top) and Program Counter (bottom) of the FFT vector implementation.} 
  \label{figFFTvehavephase}
\end{figure}

Once the vectorized implementation is ready, we instrumented the code to identify the different code phases. This step is optional, but it helps identifying the different phases later on when analyzing the trace generated when running with the Vehave emulator.
After opening the resulting trace in Paraver, we can look at the different code phases by either using the event we added in our instrumentation or looking at the Program Counter (PC).
Figure~\ref{figFFTvehavephase} shows on the $x$-axis the sequence of the 4821 vector instruction executed. The color code of the top timeline identifies the phases that we marked with the manual instrumentation. We find the highest amount of vector instructions in the code phase 2 (\ie the pink phase has the highest number of vector instruction).
The bottom plot of Figure~\ref{figFFTvehavephase} reports on the $y$-axis the value of the PC.
We can observe the saw-toothed shape of the PC, typical of an iterative execution.

\begin{figure}[htbp]
  \centering
  \includegraphics[width=.8\linewidth]{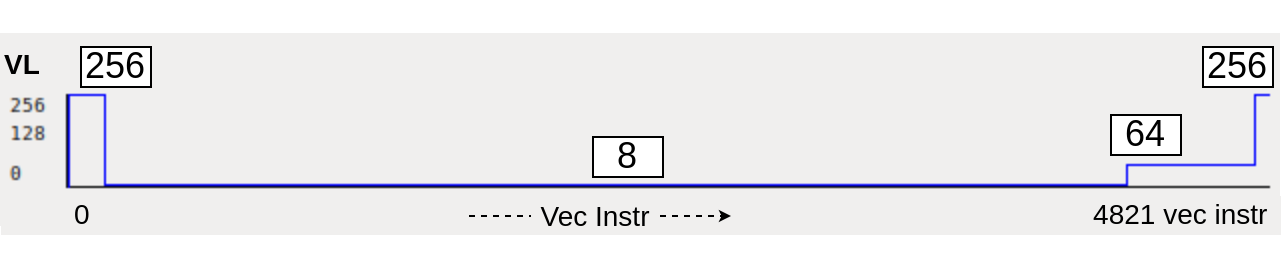}
  \caption{Vehave trace with the Vector Length per FFT phase.}
  \label{figFFTvehavevl}
\end{figure}

Since Vehave emulates each instruction individually, we also have detailed information about them, such as the number of elements used by vector instructions, the so called {\em vector length}.
With these values, we can compute the average vector length per each user-defined phase, as seen in Figure~\ref{figFFTvehavevl}.
Phase 2, where we find most instructions, also has the lowest vector-length.
Ideally, we want to use the maximum vector length of the machine (256 double precision elements) for all phases.

\begin{figure}[htbp]
  \centering
  \includegraphics[width=.8\linewidth]{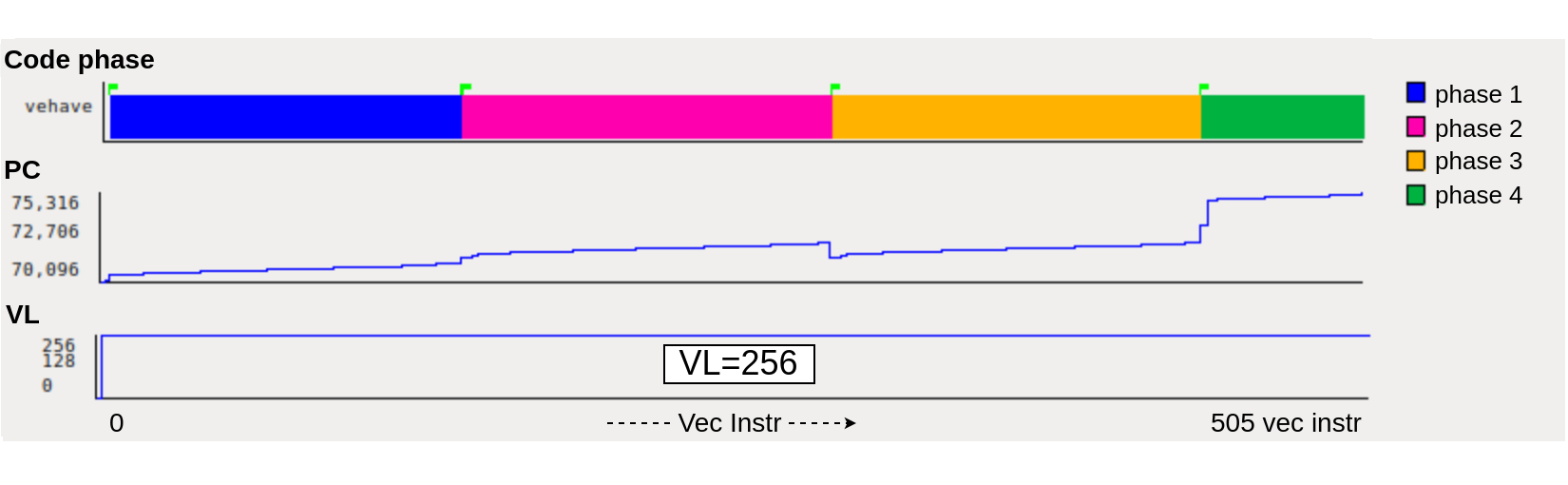}
  \caption{Vehave trace with Code phase, PC and Vector Length for the improved FFT implementation.}
  \label{figFFTvehavegather}
\end{figure}

Since we developed this first version of the vectorized FFT library, we can learn from this first observations and improve it to take advantage of the maximum vector length.
Figure~\ref{figFFTvehavegather} shows the Vehave trace of this improved implementation.

As it can be seen, all four phases have approximately the same number of instructions, phase 2 and 3 are completed with a single internal iteration, and the Vector Length is 256 for all phases.

\begin{figure}[htbp]
  \centering
  \includegraphics[width=.8\linewidth]{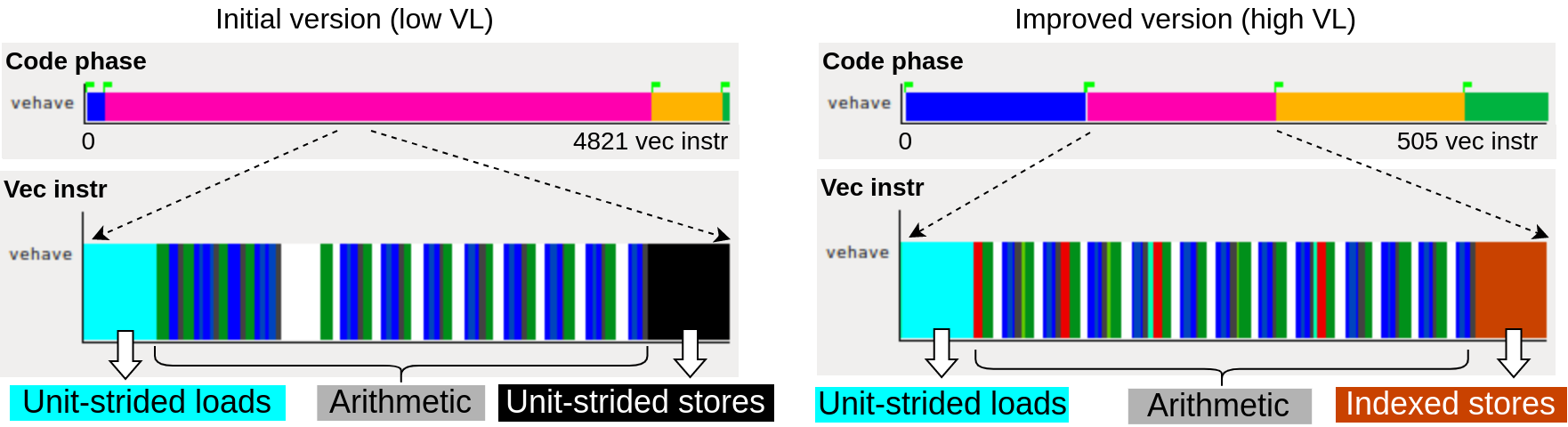}
  \caption{Vehave trace with Vector instructions in the second phase of the FFT for both implementations.}
  \label{figFFTvehaveinstr}
\end{figure}

Figure~\ref{figFFTvehaveinstr} takes a closer look into the vector instructions emulated by Vehave for both versions of the code, focusing on a single iteration of the second phase.
We see that in order to increase the vector length, the implementation now contains indexed memory operations, which have a much lower bandwidth than their unit-strided counterparts.

\subsection{Step 3: performance analysis on FPGA prototype}

\begin{figure}[htbp]
  \centering
  \includegraphics[width=.8\linewidth]{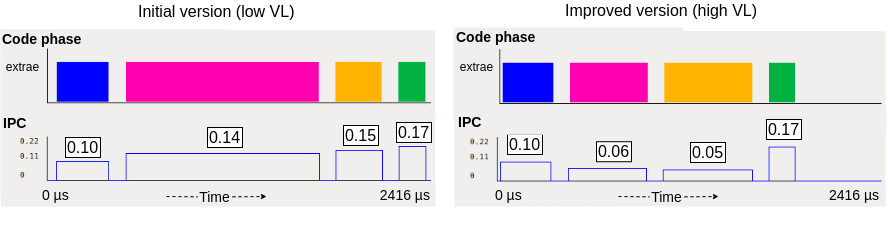}
  \caption{Extrae timelines with Phase time (top) and Instructions per Cycle (bottom) of both FFT implementations.}
  \label{figFFTextrae}
\end{figure}

After this initial evaluation using Vehave, we can jump to the FPGA system and get actual timing measurements and traces.
Figure~\ref{figFFTextrae} shows three views from Extrae traces of both implementations, obtained in the FPGA.
The duration of the pink region (phase 2) decreased when the VL changed from 8 to 256 elements. At the same time, the yellow region (phase 3) worsened when changing from 64 to 256, most likely due to the inefficient memory operations presented in Figure~\ref{figFFTvehaveinstr} outweighing the gains of a larger vector length.

Figure~\ref{figFFTextrae} shows on the $x$-axis the execution time and on the $y$-axis the value of the Instructions per Cycle (IPC) in each of the phases with vector instructions for two versions of the code.
On the left we show an execution that uses small vector lengths and on the right an execution with larger vector length.
When we change our implementation to take advantage of a larger vector length, we obtain a code that uses less instructions to process the same amount of data, reducing the overall IPC.
As a consequence, the phases affected by our optimization (pink and orange) show a reduction of IPC and the ratio of vector instructions to total instructions also decreased.

Finally, we can use the ILA in the FPGA to perform a fine-grain analysis of the vector instructions (similar to Vehave but with cycle-accurate data).

\begin{figure}[htbp]
  \centering
  \includegraphics[width=.8\linewidth]{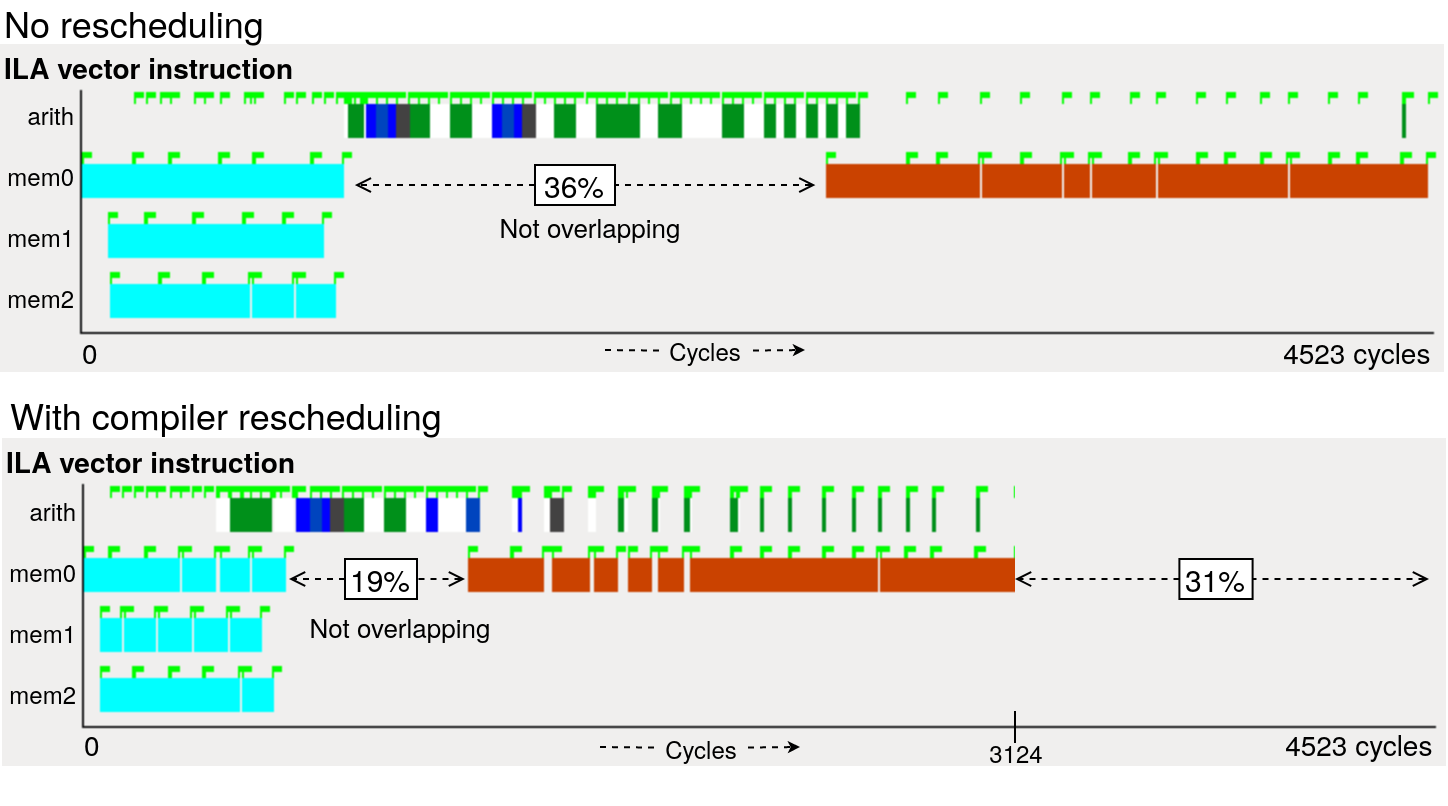}
  \caption{Vector Instructions reported in Vehave (top), in the ILA (mid), and after implementing and applying compiler rescheduling (bottom).}
  \label{figFFTila}
\end{figure}

At the top of Figure~\ref{figFFTila} we show a Vehave timeline, and below it a ILA timeline of the same region of code.
The vector instructions in the ILA timeline vary in length depending on their actual duration, and we present a different row for each hardware pipeline.
This way, we can study the parallelism between pipelines.
In this case, we detected that little to no parallelism is being exploited.

Looking at these types of traces, we gave feedback to the compiler developers so they could reschedule the instructions to overlap arithmetic and memory operations.
This is done by interleaving both types of instructions instead of grouping the loads and stores at the start and end of the iterations, respectively.
The results from this automatic rescheduling are seen at the bottom of Figure~\ref{figFFTila}.
Now, some arithmetic instructions are executed concurrently with vector loads or stores, reducing the time where there is not a memory instruction in flight.

\section{Conclusions}\label{secConclusions}

We present an incremental methodology, starting from less detailed to more detailed, which is applied to a RISC-V architecture with large vectors targeting HPC. This methodology enables the early adoption and tuning of software when developing new hardware platforms. It has the great advantage of allowing the system software to be ready before the hardware is ready. In addition, this tool allows the preparation of all layers of software required for an efficient HPC execution, such as compilers, libraries, and scientific applications.

This method enables running the same binary on commercial platforms and on the FPGA emulator. It also allows any kind of system calls, making it highly flexible for software developers. The tool infrastructure includes the ability to spy on the values of signals that are internal to the implementation, providing valuable insights. Furthermore, this methodology is cheaper than a full system simulator and runs faster than a simulator, even though it requires an RTL implementation.

The approach taken by Vehave results in a fast execution for scalar code but slow for vector instructions and it only allows to study post-mortem trace of vector instructions.
QEMU could be faster than Vehave and could allow us to gather traces of all kind of instructions, both vector and scalar. It is slower than native execution but it can run on fast x86 hosts. Indeed, we are considering it as an extension of the SDV tool-chain.
The FPGA implementation of the RTL runs at lower frequency than any ASIC, but still order of magnitude faster than any software simulator.
The study using the logic analyzer signals has great potential but it is limited by the number of instructions/signals that can be monitored.

As a future work, we plan to extend the monitored observables to include power drain and expand the toolchain to include QEMU and a multi-FPGA infrastructure.

\subsubsection*{Acknowledgments}
{ \small
This research received funding from the EU-HPC-JU under FPA N.~800928 (EPI) and SGA N.~101036168 (EPI-SGA2). The JU receives support from some member countries. The EPI-SGA2 project, PCI2022-132935 is also co-funded by MCIN/AEI /10.13039/501100011033 and by the UE NextGenerationEU/PRTR.}

\bibliographystyle{splncs04}
\bibliography{99-sdv}
%
%
%
%
\end{document}